\documentclass[prl,twocolumn,showpacs]{revtex4-1}
\usepackage{graphicx}
\usepackage{times}
\usepackage{bm}
\usepackage{amsmath, verbatim}

\def\lsim{\mathrel{\rlap{\lower4pt\hbox{\hskip1pt$\sim$}}
    \raise1pt\hbox{$<$}}}                % less than or approx. symbol
\def\gsim{\mathrel{\rlap{\lower4pt\hbox{\hskip1pt$\sim$}}
    \raise1pt\hbox{$>$}}}                % greater than or approx. symbol

\def\be{\begin{equation}}
\def\ee{\end{equation}}
\def\bea{\begin{eqnarray}}
\def\eea{\end{eqnarray}}
\def\bse{\begin{subequations}}
\def\ese{\end{subequations}}

\def\be{\begin{eqnarray}}
\def\ee{\end{eqnarray}}

\newcommand{\ket}[1]{|#1\rangle}
\newcommand{\bra}[1]{\langle #1 |}

\renewcommand{\vec}[1]{{\bf #1}}

\def\ga{{\ \lower-1.2pt\vbox{\hbox{\rlap{$>$}\lower5pt\vbox{\hbox{$\sim$}}}}\ }}
\def\la{{\ \lower-1.2pt\vbox{\hbox{\rlap{$<$}\lower5pt\vbox{\hbox{$\sim$}}}}\ }}

\def\eps{\varepsilon}

\def\beq{\begin{equation}}
\def\eeq{\end{equation}}

\def\bea{\begin{eqnarray}}
\def\eea{\end{eqnarray}}

\begin{document}

\title{Topologically non-trivial superfluid phases and Majorana fermions from Kohn-Luttinger effect}
\author{M. S. Marienko$^1$}
\author{Jay D. Sau$^2$}
\author{Sumanta Tewari$^{3}$}

\affiliation{$^1$Department of Physics and Astronomy, Hofstra University, Hempstead, New York 11549, USA\\
$^2$ Department of Physics, Harvard University, Cambridge, MA 02138\\
$^3$Department of Physics and Astronomy, Clemson University, Clemson, SC
29634}

\begin{abstract}
Spin-triplet $p$-wave superfluids of single-species (or spin-polarized) fermionic atoms
is a topological superfluid. In 2D such a superfluid supports zero-energy topological Majorana fermion excitations
in order parameter defects such as vortices and sample edges. In 3D these superfluids support topologically protected
Dirac points in the bulk spectrum and flat surface Majorana arcs. Despite the promise, creating a spin-triplet $p$-wave
single-species fermionic superfluid either from direct $p$-wave Feshbach resonance or indirectly from the recently proposed artificial
spin-orbit and Zeeman couplings can be technologically challenging. Here we show that such topological superfluids can be far
more simply created by using the
Kohn-Luttinger effect applied to two species of spin-polarized fermions with a density imbalance. We discuss how the
topological Majorana excitations of the resulting superfluids can be identified using recently-developed experimental techniques.

\end{abstract}

%\pacs{73.43.Nq,74.25.Dw,75.10.Hk}
\maketitle

%%%%%%%%%%%%%%%%%%%%%%%%%%%%%%%%%%%%%%%%%%%%%%%%%%%%%%%%%%%%%%%%%%%%%

\textit{Introduction:} Spin-triplet $p$-wave superfluids of spin-polarized fermionic atoms offer an excellent test-bed for exploring
many intriguing quantum phenomena \cite{Volovik,Read-Green,Nayak,Ivanov}. Such systems are generically topological in all spatial dimensions, exhibiting,
in $3D$, topologically protected Dirac points in the bulk spectrum \cite{Volovik,Volovik1,Zhang-Tewari} and surface Majorana arcs \cite{Sau-Tewari}, while in $D=2,1$ they exhibit zero-energy
Majorana fermions in order parameter defects such as vortices and sample edges \cite{Read-Green,Nayak}. In recent years, the physics of the 2D chiral $p$-wave (%
$p_{x}+ip_{y}$) superfluids has attracted much attention \cite{Nayak}
because of its nontrivial statistical properties \cite{Ivanov} and potential
application in topological quantum computation \cite{Nayak,Kitaev,Tewari,Zhang}. The
chiral superfluids can also act as a testbed for studying quantum
phenomena such the quantum non-locality and the violation of Bell's inequality
\cite{Zhang}, which are often masked by the many-body effects in a
macroscopic system. Schemes for observing anyonic
statistics and implementing topological quantum computation using vortices
in these systems have also been proposed \cite{Zhang}. A
potential advantage of the spin-polarized cold atom $p_{x}+ip_{y}$ chiral superfluid over
solid state chiral-$p$ wave system such as strontium ruthenate \cite{Maeno,Xia} is
that, the cold atom system being spin-polarized, even ordinary vortex excitations of the superfluid can have
 trapped Majorana modes while in the solid state system the relevant topological excitations are the exotic half-quantum vortices \cite{Ivanov,Sarma}.
  % solid state non-Abelian statistics associated with the
%Majorana fermion modes in the order parameter defects should be readily observable.

Despite the promise, creating a spin-triplet $p$-wave superfluid of single-spin (or single-species) fermionic cold atoms
has turned out to be a challenging task. With the
observations of $p$-wave Feshbach resonances in spin-polarized $^{40}$K and $%
^{6}$Li atoms in optical traps \cite{Regal,Schunck,Esslinger,Jin}, a $%
p_{x}+ip_{y}$ superfluid of fermionic atoms was thought to be soon realizable using Feshbach resonance \cite{Melo,Gurarie,Cheng}.
However, because of the short lifetimes of the $p$-wave pairs and molecules \cite{Jin}, realizing a spin-polarized $p$-wave superfluid in this route
now seems difficult. An alternative scheme, involving an artificially created spin-orbit coupling and a Zeeman field along with a
regular $s$-wave Feshbach resonance, has also been proposed \cite{Zhang-spin-orbit,Sato-Fujimoto}. Here, even though the fermions are not spin-polarized, a judicious combination
of the spin-orbit coupling, Zeeman field, and the fermion density can still be obtained to enforce a single Fermi surface, thus
allowing the formation of a topological superfluid with Majorana fermions. This strategy, which has now been theoretically applied also to condensed matter systems \cite{Sau,Annals,Alicea-Tunable,Long-PRB,Roman,Oreg}, still requires artificial laser-induced spin-orbit and Zeeman couplings for fermionic atoms which are presently beyond experimental reach.
Considering the importance of a $p$-wave superfluid of the same species of atoms both for fundamental as well as technological reasons,
it's important to investigate other, possibly simpler, methods of realizing such systems in optical traps without requiring either $p$-wave
Feshbach resonance or artificial spin-orbit and Zeeman couplings. In this paper we achieve this goal by utilizing
a completely different (and more practical in cold fermions) scheme for realizing a triplet $p$-wave superfluid of the same species of atoms by using the celebrated Kohn-Luttinger effect \cite{Kohn65,Kagan89,Chubukov93} modified to apply to the case of two types of (individually spin-polarized) fermionic atoms. The essence of the Kohn-Luttinger effect is superfluid pairing in a short-ranged \textit{repulsive} fermionic system
mediated by another fermionic background which renormalizes the bare repulsive two-particle interaction vertex. By using two distinct species of
individually spin-polarized fermions with a density imbalance we show here that it is possible to create single species $p$-wave topological superflids which can be individually probed to uncover the topological Majorana excitations.

Once a single-species spin-triplet superfluid is realized in experiments, one natural and
important question is how to experimentally observe the topological properties, in particular, the
topological excitations known as Majorana fermions. Majorana fermions are zero-energy excitations of
a 2D chiral ($p_x+ip_y$)-wave or a 1D $p_x$-wave superfluid which are trapped by order parameter defects such
as vortices (in 2D) and sample edges (in 1D). Majorana fermions, first proposed by E. Majorana to describe neutrinos \cite{Majorana}, are
quantum particles which are their own anti-particles \cite{Wilczek} (i.e., the second quantized operators $\gamma$ satisfy $\gamma^{\dagger}=\gamma$),
unlike Dirac fermions where particles and anti-particles (holes) cannot be identified. A cold-atom demonstration of Majorana
fermions in itself is a remarkable experimental goal, but also they have attracted a lot of recent attention because of their
potential use in topological quantum computation (TQC) \cite{Nayak,Kitaev}. In the condensed matter systems Majorana fermions have been proposed
to be identifiable in local STM tunneling experiments \cite{Tewari2,Long-PRB} and in fractional Josephson effect \cite{Kitaev-1D,Sengupta,Roman}.
However, such methods cannot be directly used to
detect Majorana fermions in cold atom neutral superfluids, because superfluids, in
contrast to superconductors, do not directly couple to an electric potential difference. Nevertheless, we
show how Majorana fermions in the 2D cold atom spin-triplet superfluids produced in our scheme can be identified using
radio-frequency (RF) spectroscopy experiments. In 2D, according to our theory, only the majority component (say K) of the two species of fermions
 becomes superfluid in the experimentally achievable temperatures, while the minority component (say Li) remains uncondensed. This, however, does
 not pose any problem in imaging the MFs in the majority component since in our RF spectroscopy based imaging the individual species of atoms can be
 probed independent of the other species. The absence of inter-conversion between the K and Li atoms ensures that the MFs in the majority component
 remains robust even if the minority component remains gapless and uncondensed. Note that a similar robustness of MFs to the presence of gapless electrons does not exist in condensed matter systems, and, therefore, cold atom fermionic mixtures are particularly suitable for realizing MFs which can be robustly probed using RF spectroscopy. In the 3D spin-triplet $p$-wave superfluids produced in our scheme the topological
bulk Dirac modes have a structure similar to that in the recently proposed spin-orbit coupled 3D superfluid in Ref.~[\onlinecite{Zhang-Tewari}]. We discuss how these non-trivial excitations can be probed in the cold atom analog of the angle-resolved
photoemission experiments. Taken together, the method to create a single-species spin-triplet $p$-wave
superfluid using the Kohn-Luttinger effect, coupled with the methods
to observe the non-trivial topological excitations, gives a complete
description of a promising new way to create and analyze a topologically non-trivial
superfluid in both two and three dimensions.

\textit{Kohn-Luttinger effect and superfluid critical temperatures:}
The essence of the Kohn-Luttinger effect is the possibility of a superfluid pairing in fermionic systems with short-range repulsion that does not support the conventional BCS-type pairing \cite{Kohn65}. The presence of a fermionic background renormalizes the two-particle vertex, and its angular dependence in momentum space (Kohn singularity) results in the long-range Friedel oscillations of the two-particle potential in real space, inverting the sigh of the effective interaction. A diagrammatic expansion of the effective vertex at low fermionic density is known to result in the effective attraction in the second order of perturbation theory in 3D (third order in 2D) \cite{Kagan89, Chubukov93}.

\begin{figure}[t]
     \centerline{   \includegraphics[width=5 cm]{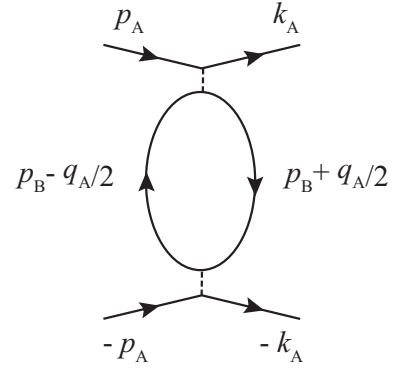}}
    \caption{Effective interaction between fermions of the type A mediated by fermions of the type B in second order perturbation theory.}
    \label{fig:_interaction}
\end{figure}

We apply this idea to the mixture of two types of spin-polarized cold gases with different masses and densities. Such two-component fermion systems have been realized either as mixtures of different atomic states of $^6$ Li \cite{thomas} and $^{40}$ K \cite{demarco} and also mixtures of $^6$Li-$^{40}$K \cite{wille}. For short-range interaction potential, the antisymmetry of the fermionic wave function results in the absence of diagonal intra-component terms in the interaction Hamiltonian. The leading repulsive bare interaction is reduced to the isotropic $s$-wave scattering between atoms of different kind with the scattering amplitude $a_0 >0$. Such a system can be described by the following Hamiltonian:
\bea\label{Hamiltonian_Fourier_transformed}
\hat{H}' & = & \sum_{i=A, B}\sum_{\vec{p}} \eps_i (p) c^\dagger_{\vec{p}i} c_{\vec{p} i}\\ &+& U_{AB} \sum_{i\neq j}\sum_{\vec{p}\vec{p}'\vec{q}} c^\dagger_{\vec{p}i}c^\dagger_{\vec{p}'+\vec{q}j}c_{\vec{p}+\vec{q}j}c_{\vec{p}'i} ,\nonumber
\eea
where $i$ and $j$ indicate species of two different types (A and B). The effective interaction  $\tilde{\Gamma}_{AA}$ between fermions of the type A in the second order of perturbation theory (the lowest order where the effect exists) is mediated by the fermions of the type B and is represented diagrammatically in the Fig. \ref{fig:_interaction}.
In the low-density limit $r_0 p_F \ll 1$ where $r_0$ is the range of the potential, the effective interaction term Fig. \ref{fig:_interaction} becomes $\tilde{\Gamma}_{AA}(q_A)=-\lambda^2 \Pi_{BB}(q_A)$,
where $q_A = p_{FA}\sqrt{2(1-\cos\theta)}$ is the difference between the incoming and outgoing momenta $\vec{p}_A$ and $\vec{k}_A$ with $\theta$ the angle  between them, $\lambda = 2 a_0 \sqrt{p_{FA}p_{FB}}/\pi$ is the modified 3D gas parameter in the case of inter-component scattering. Furthermore, $\Pi_{BB}(q_A)\sim \int \frac{d^4 p_B}{(2 \pi)^4} G\left(p_B+\frac{q_A}{2}\right)G\left(p_B-\frac{q_A}{2}\right)$ is the polarization operator that corresponds to the Green's function loop in the Fig. \ref{fig:_interaction}. Integrating $\tilde{\Gamma}_{AA}(q_A)$ at the Fermi surface with eigenfunctions of the angular momentum (Legendre polynomials $P_l(\cos(\theta))$ in 3D), one can see that the most negative eigenvalue $\tilde{\Gamma}^{(l)}_{AA}=\int \tilde{\Gamma}_{AA}(q_A(\theta)) P_l(\cos(\theta)) \frac{d\cos\theta}{2}$ corresponds to $l=1$ and the triplet superfluid state is formed in the system A. The critical temperature is then given by:
\beq\label{Tc_general}
T^{(1)}_{cA}\sim \eps_{FA} \exp\left[ -1/\left( N_{FA} |\tilde{\Gamma}^{(1)}_{AA}| \right)\right],
\eeq
where $N_{Fi}$ and $\eps_{Fi}$ are the density of states at the Fermi surface and the Fermi energy of the species $i$.

\begin{figure}[t]
     \centerline{   \includegraphics[width=8 cm]{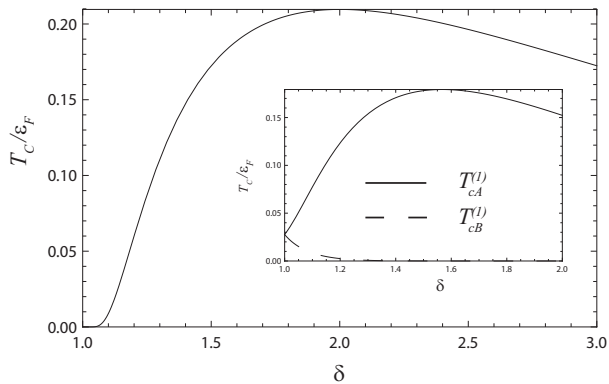}}
    \caption{Triplet critical temperature of the majority species A as a function of population imbalance $\delta = p_{FA}/p_{FB}$, at $p_{FA}>p_{FB}$ and $f_0 \sim 0.8$ in 2D.  The minority component B remains in the normal state in the second order of perturbation theory. Inset: The critical temperatures for the species A and B in 3D at $\lambda \sim 1.9$. }
    \label{fig:_Tcs}
\end{figure}

The static polirization operator in 3D is given by:
\bea\label{Pi_BB_3D}
\! \Pi_{BB}(q_A) = \frac{m_B p_{FB}}{4\pi}\left[ 1+\frac{4 p_{FB}^2 -q_A^2}{2p_{FB} q_A} \ln \frac{2p_{FB}+q_A}{|2p_{FB}-q_A|} \right]
\eea
then the critical temperature $T^{(1)}_{cA}$ is
\beq\label{TcA}
T^{(1)}_{cA}\sim \eps_{FA} \exp\left[ -1/\left(\lambda^2 f(\delta)\right)\right],
\eeq
where
\beq\label{f_delta}
f (\delta)=\frac{4\pi}{m_B p_{FB}} \int_{-1}^1 \Pi_{BB}(q_A) P_1 (\cos\theta) \frac{d\cos\theta}{2}
\eeq
is a dimensionless function of the population imbalance $\delta = p_{FA}/p_{FB}$ \cite{Kagan89, Baranov92}. One can see that $f(\delta)$, and consequently $T^{(1)}_{cA}(\delta)$ increases at small imbalance  $\delta  \ga 1$, has a maximum, and then drops to zero at $\delta\gg 1$. Two factors contribute to the effect: on one hand, Kohn singularity increases with $\delta$ and this enhances the effective interaction of the fermions of the type $A$; on the other hand, the density of the species $B$ decreases as $\delta$ increases, and this eventually results in the total absence of pairing at $\delta \rightarrow \infty$. The effective interaction for the species $B$ is, in turn, mediated by the subsystem $A$. To obtain the critical $p$-wave temperature $T^{(1)}_{cB}$ one should swap indices A and B in the Eqs. (\ref{Pi_BB_3D}) - (\ref{f_delta}). Then, the temperature $T^{(1)}_{cB}$ decreases monotonically as $\delta$ increases. At $\delta=1$ the system is equivalent to the non-polarized 3D Fermi gas, and the superfluidity occurs in both subsystems simultaneously at $T^{(1)}_c \sim \eps_F \exp\left[ -13/\lambda^2 \right]$ \cite{Kagan89}. Thus, at a finite imbalance $\delta > 1$ there is a temperature window $T^{(1)}_{cB}<T <T^{(1)}_{cA}$ where only the component $A$ is superfluid, while the other component $B$ remains in the normal state (See Fig. \ref{fig:_Tcs}).

 Similar arguments apply to the 2D case, where $\cos l\varphi$, the 2D eigenfunctions of the angular momentum, replace the Legendre polynomials, and $q_A = p_{FA}\sqrt{2(1-\cos\varphi)}$. The polarization operator
\beq\label{2D_Pi}
\Pi_{BB}(q_A)= \frac{m_B}{2\pi} \left(1-{\mbox Re} \sqrt{1-4p_{FB}^2/q_A^2} \right)
\eeq
explicitly depends on $\varphi$ (and thus on $q_A$) only if $p_{FA}>p_{FB}$. In the opposite case, $p_{FA}\leq p_{FB}$, the angular dependence is completely excluded from the real part of the Eq. (\ref{2D_Pi}) and all the eigenvalues $\tilde{\Gamma}^{(1)}_{AA}$ vanish. The population imbalance becomes the necessary condition for the pairing to exist in the second order of perturbation theory. The $p$-wave superfluid state is then formed by the species with larger value of the Fermi momentum (subsystem A), with the critical temperature given by the equation \cite{Kagan89}:
\beq\label{TcA_2D}
T^{(1)}_{cA}\sim \eps_{FA} \exp\left[ - \delta^2  / (4(\delta-1)f_0^2 )\right],
\eeq
where $f_0$ is the 2D gas parameter that, as usual, is obtained by the renormalization of the two-particle vertex and that in 2D contains a logarithm $f_0 \approx -1/2 \ln r_0 p_F$.  Just like the 3D case, the critical temperature (\ref{TcA_2D}) has a maximum at some $\delta \approx 2 $ (See Fig. \ref{fig:_Tcs}). The other, low-density subsystem B remains in the normal state at all temperatures in this order.
Note that, from Eq.~(\ref{TcA_2D}), $T^{(1)}_{c}=0$ at $\delta=1$ in the second order of perturbation theory, and one needs to include the next, third-order terms  of the expansion. To this order, the critical temperatures for both subsystems (note that for $\delta=1$ the components A and B should
 behave similarly) are given by $T^{(1)}_{c} \sim \eps_F \exp\left[ -1/(6.1 f_0^3) \right]$ \cite{Chubukov93}.

 Note that, according to Eq.~(\ref{TcA_2D}) and Fig. \ref{fig:_Tcs}, at the gas parameter $f_0 \sim 0.8$, the 2D maximum $T_{cA}^{(1)} \sim 0.2 \epsilon_F$ is well inside the currently experimentally achievable range \cite{Gaebler}. The 3D critical temperature is within the experimental range at $\lambda \sim 1.9$ in Fig. \ref{fig:_Tcs}. The applicability of the perturbation theory at these values of the gas parameter requires further justification, in particular, higher order terms in the perturbation theory which we leave for future work. Here we, however, remind that, as was demonstrated in Ref. \cite{Kagan89}, the zero-imbalance critical temperature at the gas parameter $\lambda \sim 1.3$ (same as that in the superfluid $^3$He) has the right order of magnitude $T_{c}^{(1)} \sim 10^{-3} \epsilon_F$.

In the present proposal we ignore possible off-diagonal exchange terms in the Hamiltonian (\ref{Hamiltonian_Fourier_transformed}) that scatter Cooper pairs between different bands in the two-band model \cite{Moskalenko - Suhl}. These terms will create a superfluid state in both bands simultaneously \cite{Baranov92}. In the present case where components A and B represent two types of atoms (e.g., $^{6}$Li and $^{40}$K) such off-diagonal terms
 cannot exist. In addition, we see that when both species are completely polarized, the critical temperatures splitting is mostly affected by the population imbalance, or the ratio of Fermi momenta of different components, but not the by the ratio of their masses.

\begin{comment}
 The proposed method can be extended onto a multi-component system. The case of several hyperfine components of the same atoms in a trap was considered in the Ref. \cite{Baranov96}. It was shown that the effective interaction in the component with the highest value of Fermi momentum is mediated by the entire fermionic background. The effective vertex in the component A is then sum of polarization operators in components B, C, etc.: $\tilde{\Gamma}_{AA}(q_A)=- U_{AB}^2 \Pi_{BB}(q_A)- U_{AC}^2 \Pi_{CC}(q_A)+\ldots$.

  Ultracold alkali gases in traps have densities between $10^{12}$ cm$^{-3}$ to $10^{15}$ cm$^{-3}$ and they can be cooled
 down to temperatures from $10^{-6}$ K to $10^{-8}$ K. A simple estimation for the Fermi-energy in 3D
 $\eps_F=\frac{\hbar^2}{2m} \left( 3 \pi^2 n \right)^{2/3}$  gives $\eps_F\sim 10^{-7} - 10^{-6}$ K for both
 $^6$Li and $^{40}$K. Thus the critical temperature in the extreme case when the expansion parameter of the
 perturbation theory (the gas parameter) $\lambda\la 1$ is of the order of $T_C \sim 10^{-8} - 10^{-7}$ K.

\end{comment}

\textit{RF spectroscopy based detection of Majorana fermions:}
Topological superfluids of the type considered here are characterized in $D=2$ by Majorana fermion excitations that are localized in defects of the  superfluid order parameter.
Such zero-energy states can be detected as resonances in tunneling near zero energy, which can be realized
in cold atomic system using RF spectroscopy \cite{Tewari,regaljin,sgupta,chin}.
%Following the discussion in Ref.~\cite{sumantazoller}, we review how a localized MF
%mode may be detected using RF spectroscopy.
Let us consider a two-component mixture of spin-polarized $^{40}$K as the majority species A and $^6$Li atoms as the minority species
B. We first note that for the purpose of imaging MFs using RF spectroscopy, the fermions in the components A and B can be discussed independently because of the absence of inter-conversion between the two species of atoms. As discussed above, from the Kohn-Luttinger effect, the majority A atoms can be expected to be in a topological superfluid
phase, while, in $D=2$ the minority component B remains uncondensed. The uncondensed B atoms, however, pose no problem for the detection
 of the MFs in component A as we clarify later. Suppose the $^{40}$K atoms in the superfluid are in the $4^2S_{1/2}$
hyperfine ground state, $\ket{i}=\ket{F = 9/2,m_F = -7/2}$.
In this case, the application of a two-photon Raman pulse can transfer
the $^{40}$K atoms from the state $\ket{i}$ to another hyperfine state $\ket{j}=\ket{F = 7/2,m_F = -5/2}$.
Within the rotating wave-approximation (RWA), the frequency difference between
the two Raman lasers $\Omega$ can be considered to be the detuning between the energy levels of the
internal states $\ket{i}$ and $\ket{j}$.
The effective Hamiltonian describing such a pair of Raman beams within the RWA is written as
\begin{equation}
H_{Raman}=\int d\bm r \{f(\bm r)[\psi_i^\dagger(\bm r)\psi_j(\bm r)+h.c]+\Omega \psi_j^\dagger(\bm r)\psi_j(\bm r)\},
\end{equation}
where $\psi_{i,j}^\dagger(\bm r)$ are the creation operators for atoms in states $\ket{i}$ and $\ket{j}$ respectively.
Here we assume that a localized Raman beam with a  beam waist width $w\approx 1.5\, \mu$m is much smaller
than the typical distance $10\mu m$ between vortices \cite{zwierlein} is being used. The function $f(\bm r)$ is localized
within the beam waist.

The above Hamiltonian $H_{Raman}$ transfers atoms from the superfluid of $^{40}$K atoms in the $\ket{i}$ state to
atoms in the $\ket{j}$ state, which can be thought of as being in an empty reservoir. This is similar to a tunneling
probe in a superconductor, where electrons can tunnel from the superconductor into the leads. For topological superfluids without
  vortices and other defects, which have a unique ground state $\ket{\Psi}$ together with a finite quasiparticle
gap $\Delta_0$ to the lowest excited state $\ket{\Psi_1}$, the rate of generation of atoms in state $\ket{j}$ must vanish
for energies $\Omega<E_F-\Delta_0$ (where $E_F$ is the fermi-energy of the superfluid of $\ket{i}$ atoms)
 because of energy conservation.
In contrast, topological superfluids with vortices contain zero-energy MF excitations bound to vortices.
The MFs are associated with self-hermitean creation fermionic operators $\hat{\gamma}$,
which commutes with the total mean-field Hamiltonian for the topological
superfluid. As a result if $\ket{\Psi}$ is a lowest-energy
state of the Hamiltonian, $\ket{\Psi_1}=\hat{\gamma}\ket{\Psi}$ is another state with exactly the same
energy (measured relative to the fermi-energy).
 Therefore the ground-state is degenerate in this case and it is possible to generate particles in the state $\ket{j}$ even
when the Raman beams are tuned to be on resonance at $\Omega=E_F$ . In fact, the rate of generation of fermions in the state $\ket{j}$
will have a peak at $\Omega=E_F$ given by
\begin{equation}
R=\sum_n |\bra{\Psi_n}H_{Raman}\ket{\Psi}|^2\delta (E-E_n-\Omega-E_F)\propto \delta(\Omega)
\end{equation}
where $\ket{\Psi_n}$ are excited states with energy $E_n$ and $E$ is the energy of the ground state $\ket{\Psi}$.
 Note that, while the absorption of a Raman beam transfers the
many-body state from $\ket{\Psi}$ to $\ket{\Psi_1}$ in principle, this does not lead to any observable time-dependence
of the particle-generation rate $R$ since the matrix elements $H_{Raman}$ for localized Raman beams
 are identical for the states $\ket{\Psi}$ and $\ket{\Psi_1}$. Therefore, the rate of particle-generation $R$ is independent
of which initial topologically degenerate state $\ket{\Psi}$ or $\ket{\Psi_1}$ one starts with, consistent with the idea
of local non-observability of the topological degree of freedom.
The tunneling operation induced by the Raman-pulse cannot be used to read-out the MF degree of freedom unless the two vortices
are brought together to induce a splitting - effectively fusing the vortices.

Another interesting feature of using a mixture of fermions
containing two different species of fermions is the topological robustness despite the presence of an even number of fermi
surfaces.
This is typically a problem in spinful chiral $p$-wave superconductors such as SrRuO$_4$ where a vortex contains a spinful Majorana
fermion with a spin-degeneracy. In that case, transverse magnetic fields which mix the two spin-species would split the Majorana fermions
and lift the topological degeneracy. The observation of true MFs in this system requires
 the use of half-quantum vortices \cite{Ivanov,Tewari-Half}, which are effectively vortices in one of the spin
species to trap Majorana fermions without a spin degeneracy. The cold atomic system proposed here provides a unique
 robustness in this regard, because here the two species
or pseudo-spins are really two different elements Li and K in this case, so that the Majorana fermions are protected
by the additional number conservation symmetry in this system. Therefore, even if one used a regular vortex, which
 produced a phase winding in each of the species separately,
there would be a pair of Majorana fermions in each vortex, which cannot be split simply because it is not possible to convert
an atom of one element to another. A similar protection would continue to exist if one of the species (say Li) was not condensed, as is the case in 2D (Fig.~2).
Conventionally, one could worry that the thermal Li gas serves as a fermion reservoir which is known to destroy the fermion
parity protection of Majorana fermions. However, since Li atoms cannot convert into K atoms, the Majorana fermions associated
with the K topological superfluid remain robust.

In $D=3$, the topological superfluids considered here are characterized by topologically protected gapless Dirac points in the bulk (Weyl fermions)
and flat Majorana modes on the surface \cite{Volovik1,Zhang-Tewari,Sau-Tewari}. Such momentum resolved gapless spectrum can in principle be detected by recently developed momentum resolved photo-emission spectroscopy for cold atoms \cite{Stewart,Gaebler}.

\textit{Conclusion:} Using Kohn-Luttinger effect we show that in a mixture of two species of individually spin-polarized fermions with population imbalance it is possible to create topologically non-trivial superfluid states characterized by Majorana fermion excitations at order parameter defects.
The Kohn-Luttinger critical temperatures of the individual superfluids are split as a function of the population imbalance and are well inside
experimentally achievable range in current experiments. The Majorana fermions in the individual components of the superfluid can be detected by RF spectroscopy. Because of the absence of inter-conversion of the different types of atoms, the Majorana fermions are detectable even if one of the components remain uncondensed.  Our proposal to create topological spin-triplet $p$-wave
superfluids using the Kohn-Luttinger effect, coupled with the methods
to observe the non-trivial topological excitations, gives a complete
description of a promising new way to create and analyze a topologically non-trivial
superfluid in both two and three dimensions.

\textit{Acknowledgements:}
We would like to thank M. Yu. Kagan for helpful discussions. S.T. thanks DARPA and NSF for support. J.S. thanks the Harvard Quantum Optics Center for support.

%%%%%%%%%%%%%%%%%%%%%%%%%%%%%%%%%%%%%%%%%%%%%%%%%%%%%%%%%%%%%%%%%%%%

%%%%%%%%%%%%%%%%%%%%%%%%%%%%%%%%%%%%%%%%%%%%%%%%%%%%%%%%%%%%%%%%%%%%%
%%%%%%%%%%%%%%%%%%%%%%%%%%%%%%%%%%%%%%%%%%%%%%%%%%%%%%%%%%%%%%%%%%%%%

%\bibliography{./refs_pwave_top_ins}

\begin{thebibliography}
%{Bangura et~al.(2008)Bangura, Fletcher, Carrington,
%Levallois, Nardone, Vignolle, Heard, Doiron-Leyraud, LeBoeuf, Taillefer
%et~al.}
\bibitem{} \expandafter\ifx\csname natexlab\endcsname\relax

\fi
\expandafter\ifx\csname bibnamefont\endcsname\relax

\fi
\expandafter\ifx\csname bibfnamefont\endcsname\relax

\fi
\expandafter\ifx\csname citenamefont\endcsname\relax

\fi
\expandafter\ifx\csname url\endcsname\relax

\fi
\expandafter\ifx\csname urlprefix\endcsname\relax

\fi
\providecommand{\bibinfo}[2]{#2} \providecommand{\eprint}[2][]{\url{#2}}

\bibitem{Volovik} G. E. Volovik, \textit{The Universe in a Helium Droplet}
(Clarendon Press, Oxford, 2003).

\bibitem{Read-Green} N. Read and D. Green, Phys. Rev. B \textbf{61}, 10267
(2000).

\bibitem{Nayak} C. Nayak \textit{et al.}, Rev. Mod. Phys. \textbf{80},
1083 (2008).


\bibitem{Ivanov} D. A. Ivanov, Phys. Rev. Lett. \textbf{86}, 268 (2001).

\bibitem{Volovik1}T.T. Heikkila, N.B. Kopnin, G.E. Volovik, Pis'ma ZhETF \textbf{94}, 252 (2011).

\bibitem{Zhang-Tewari}M. Gong, S. Tewari, C. W. Zhang, Phys. Rev. Lett. \textbf{107}, 195303 (2011).



\bibitem{Sau-Tewari} J. D. Sau and S. Tewari, arXiv: 1110.4110.

\bibitem{Kitaev} A. Kitaev, Ann. Phys. \textbf{303}, 2 (2003).

\bibitem{Tewari} S. Tewari, S. Das Sarma, C. Nayak, C. W. Zhang, P. Zoller, Phys. Rev. Lett. \textbf{98},
010506 (2007).

\bibitem{Zhang} C. W. Zhang, S. Tewari, S. Das Sarma, Phys. Rev. Lett. \textbf{99},
220502 (2007).



\bibitem{Maeno} A. P. Mackenzie and Y. Maeno, Rev. Mod. Phys. \textbf{75}, 657
(2003).

\bibitem{Xia} J. Xia \textit{et al.}, Phys. Rev. Lett. \textbf{97}, 167002
(2006).

\bibitem{Sarma} S. Das Sarma, C. Nayak, S. Tewari, Phys. Rev. B \textbf{73},
220502(R) (2006).

\bibitem{Regal} C. A. Regal \textit{et al.}, Phys. Rev. Lett. \textbf{90},
053201 (2003).

\bibitem{Schunck} C. H. Schunck \textit{et al.}, Phys. Rev. A \textbf{71},
045601 (2005).

\bibitem{Esslinger} K. G\"{u}nter \textit{et al.}, Phys. Rev. Lett. \textbf{%
95}, 230401 (2005).

\bibitem{Jin} J.P. Gaebler \textit{et al.}, Phys. Rev. Lett. \textbf{98},
200403 (2007).

\bibitem{Melo} S. S. Botelho and C. A. R. Sa de Melo, J. Low Temp. Phys.
\textbf{140}, 409 (2005).

\bibitem{Gurarie} V. Gurarie \textit{et al.}, Phys. Rev. Lett. \textbf{94},
230403 (2005).

\bibitem{Cheng} C.-H. Cheng and S.-K. Yip, Phys. Rev. Lett. \textbf{95},
070404 (2005).

\bibitem{Zhang-spin-orbit}C. W. Zhang, S. Tewari, R. M. Lutchyn, S. Das Sarma, Phys. Rev. Lett. \textbf{101}, 160401 (2008).

\bibitem{Sato-Fujimoto} M. Sato, Y. Takahashi, S. Fujimoto, Phys. Rev. Lett. \textbf{103}, 020401 (2009).

\bibitem{Sau} Jay D. Sau, R. M. Lutchyn, S. Tewari, S. Das Sarma,
 Phys. Rev. Lett. \textbf{104}, 040502 (2010).

\bibitem{Annals}S. Tewari, J. D. Sau, S. Das Sarma, Annals Phys. (N.Y.) \textbf{325}, 219, (2010).

\bibitem{Alicea-Tunable} J. Alicea, Phys. Rev. B \textbf{81}, 125318 (2010).


 \bibitem{Long-PRB} J. D. Sau, S. Tewari, R. Lutchyn, T. Stanescu and S. Das Sarma, Phys. Rev. B \textbf{82}, 214509 (2010).






 %\bibitem{unpublished} S. Tewari, J. D. Sau, S. Das Sarma, unpublished (2009).



 \bibitem{Roman} R. M. Lutchyn, Jay D. Sau, S. Das Sarma, Phys. Rev. Lett. \textbf{105}, 077001 (2010) .

\bibitem{Oreg} Y. Oreg, G. Refael, F. V. Oppen, Phys. Rev. Lett. \textbf{105}, 177002 (2010).



\bibitem{Kohn65}  W. Kohn, J.M. Luttinger, Phys. Rev. Lett \textbf{15}, 524 (1965).


\bibitem{Kagan89} M.Yu. Kagan, A.V. Chubukov, JETP Lett. \textbf{47}, 525
(1988); M.Yu. Kagan, A.V. Chubukov, JETP Lett. \textbf{50}, 517 (1989).



\bibitem{Chubukov93} A.V. Chubukov, Phys. Rev. B \textbf{48}, 1097 (1993), D.V. Efremov, M.S. Mar'enko, M.A. Baranov, M.Yu. Kagan, Physica B \textbf{284-288}, 210 (2000); M.Yu. Kagan, D.V. Efremov, M.S. Marienko, and V.V. Valkov,  JETP Letters \textbf{93}, 807 (2011).


\bibitem{Majorana} E. Majorana, Nuovo Cimento \textbf{5}, 171 (1937).

\bibitem{Wilczek}F. Wilczek, Nature Physics \textbf{5}, 614 (2009).

\bibitem{Tewari2} S. Tewari, C. W. Zhang, S. Das Sarma, C. Nayak, D.-H. Lee, Phys. Rev. Lett. \textbf{100},
027001 (2008).

\bibitem{Kitaev-1D} A. Yu. Kitaev, Phys. Usp.\textit{\ }\textbf{44} (suppl.),
131 (2001).

\bibitem{Sengupta}H.-J Kwon, K. Sengupta, V. Yakovenko, European Physical Journal B \textbf{37}, 349-361 (2004).

\bibitem{thomas}K. M. O. Hara, S. L. Hemmer, M. E. Gehm, S. R. Granade, and
J. E. Thomas, Science \textbf{298}, 2179 (2002).

\bibitem{demarco}B. DeMarco and D. S. Jin, Science \textbf{285}, 1703 (1999).

\bibitem{wille}E. Wille et al., Phys. Rev. Lett. \textbf{100}, 053201 (2008).

\bibitem{Baranov92} M.A. Baranov, M.Yu. Kagan, JETP \textbf{75}, 165 (1992).

\bibitem{Gaebler} J. P. Gaebler, J. T. Stewart, T. E. Drake, D. S. Jin1, A. Perali, P. Pieri and G. C. Strinati, Nature Physics \textbf{6}, 569 (2010).

\bibitem{Moskalenko - Suhl} V. Moskalenko, Fiz. Met. Metalloved. \textbf{8}, 503 (1959); H. Suhl, B. Mattias,
and L. Walker, Phys. Rev. Lett. \textbf{3}, 552 (1959).

%\bibitem{Baranov96} M.A. Baranov, Yu. Kagan, and M.Yu. Kagan, JETP Lett. \textbf{64}, 301 (1996).

\bibitem{regaljin}C. A. Regal and D. S. Jin, Phys. Rev. Lett. \textbf{90}, 230404 (2003).
\bibitem{sgupta} S. Gupta et al., Science \textbf{300}, 1723 (2003).
\bibitem{chin} C. Chin et al., Science \textbf{305}, 1128 (2004).
%\bibitem{sumantazoller} S. Tewari et al., Phys. Rev. Lett. \textbf{98}, 010506 (2007).

\bibitem{zwierlein}M.W. Zwierlein, et al., Nature \textbf{435}, 1047 (2005).

\bibitem{Ivanov} D. A. Ivanov, Phys. Rev. Lett. \textit{86}, 268 (2001).

\bibitem{Tewari-Half} S. Das Sarma, C. Nayak, S. Tewari, Phys. Rev. B \textit{73}, 220502 (R) (2006). 

\bibitem{Stewart}J. T. Stewart \textit{et al.,} Nature \textbf{454}, 744 (2008).

%\bibitem{Gaebler} J. P. Gaebler \textit{et al.,} Nature Physics \textbf{6}, 569 (2010).
\end{thebibliography}

%\end{document}

%%%%%%%%%%%%%%%%%%%%%%%%%%%%%%%%%%%%%%%%%%%%%%%%%%%%%%%%%%%%%%%

\end{document}